\documentclass[10pt,usletter,twocolumn]{article}
\usepackage[top=2cm, bottom=2cm, left=1.8cm, right=1.8cm]{geometry}

\usepackage{antpolt} %
\usepackage[QX]{fontenc}

\usepackage{amsmath,amssymb,amsfonts}
\usepackage{listings}                       %
\usepackage[noend]{algpseudocode}
\usepackage{multirow}
\usepackage{graphicx}		%
	\setkeys{Gin}{width=\textwidth, height=\textheight, keepaspectratio}
\usepackage{pifont}
\usepackage{textcomp}
\usepackage{xcolor}
\usepackage{algorithm,algorithmicx,algpseudocode,verbatim,color}
\makeatletter
\algnewcommand{\LineComment}[1]{\Statex \hskip\ALG@thistlm \(\triangleright\) #1}
\makeatother
\makeatletter
\algnewcommand{\BlankLine}[1]{\Statex \hskip\ALG@thistlm #1}
\makeatother
\usepackage{titling}
\usepackage[small,bf]{caption}
\usepackage[square,sort,comma,numbers]{natbib}

\usepackage{xcolor}
\usepackage{booktabs}
\usepackage{paralist}
\usepackage{threeparttable}

\usepackage{titlesec}
\titlespacing*{\section}{0pt}{*3}{6pt}
\titlespacing{\subsection}{0pt}{*2.5}{6pt}
\titlespacing{\subsubsection}{0pt}{*2}{2pt}

\definecolor{providercol}{HTML}{2b8a3e}
\definecolor{adversarycol}{HTML}{c92a2a}
\definecolor{linkcol}{rgb}{0,0,0.5}
\definecolor{citecol}{rgb}{0,0.5,0.3}
\definecolor{urlcol}{rgb}{0.3,0,0}

\usepackage{xspace}

\graphicspath{{figures/}}

\renewenvironment{thebibliography}[1]{
  \begin{oldthebibliography}{#1}
    \setlength{\itemsep}{0.0em}
    \setlength{\parskip}{0.0em}
}
{
  \end{oldthebibliography}
}

\usepackage[hang,flushmargin]{footmisc}

\renewcommand{\footnoterule}{%
  \kern -3pt
  \hrule width 1in
  \kern 2pt
}

\algrenewcomment[1]{\(\triangleright\) #1}

\usepackage[hyphens]{url}

\makeatletter
\def\url@leostyle{%
  \@ifundefined{selectfont}{\def\UrlFont{}}%
  {\def\UrlFont{}}%
}
\makeatother
\usepackage[hyphenbreaks]{breakurl}

\definecolor{darkred}{RGB}{153,0,0}
\definecolor{darkblue}{RGB}{0,0,99}
\usepackage[colorlinks=true, linkcolor=darkred, citecolor=darkred, urlcolor=darkblue]{hyperref}

\usepackage[aboveskip=2pt,belowskip=0pt]{subcaption}
\usepackage{indentfirst}

\usepackage{amsmath,amsthm,amssymb,graphicx}
\usepackage{algorithm,algorithmicx,algpseudocode,verbatim,color}
\usepackage{enumitem}
\usepackage[all]{xy}
\usepackage{flushend}

\usepackage{tabularx}
\usepackage{textcomp}
\usepackage{xcolor}
\usepackage{booktabs}

\newif\ifcomment
\commenttrue 

\newcommand{\myparagraph}[1]{\smallskip\noindent\textbf{#1}}

\ifcomment
	\newcommand{\edc}[1]{\textbf{\em\color{brown}EDC: #1}}
	\newcommand{\parth}[1]{\textbf{\em\color{red}PB: #1}}
\else
	\newcommand{\edc}[1]{}
	\newcommand{\parth}[1]{}
\fi

\begin{document}

\setcounter{secnumdepth}{2}

\title{\bf Group-Differentiated Discourse on Generative AI in High School Education: A Case Study of Reddit Communities}

\author{Parth Gaba$^1$ and Emiliano De Cristofaro$^1$\\[1ex]
\normalsize $^1$Valley Christian  High School, San Jose \;\;\;\;  $^2$University of California, Riverside}
\date{}

\maketitle

\begin{abstract}
In this paper, we study how different Reddit communities discuss generative AI in high school education, focusing on learning, academic integrity, AI detection, and emotional framing. Using 3,789 posts from five education-related subreddits, we compare student, teacher, and mixed communities using a pipeline that combines keyword retrieval, human-validated relevance filtering, LLM-assisted annotation, and statistical tests of group differences. 

We find that stakeholder position strongly shapes discourse: teachers are more likely to articulate explicit pedagogical trade-offs, simultaneously framing AI as both beneficial and harmful for learning, whereas students more often discuss AI tactically in relation to accusations, grades, and enforcement. Across all groups, detector-related discourse is associated with significantly higher negative emotion, with larger effects for students and mixed communities than for teachers. These results suggest that AI detectors function not only as contested technical tools but also as governance mechanisms that impose asymmetric emotional burdens on those subject to institutional enforcement. Finally, we argue that detection-based enforcement should not serve as a primary academic-integrity strategy and that process-based assessment offers a fairer alternative for verifying authorship in AI-mediated classrooms.
\end{abstract}

\section{Introduction}

In high school classrooms, AI tools are being increasingly used for a variety of reasons, ranging from personalized tutoring~\cite{kasneci2023} to learning support~\cite{mollick2023} and beyond.
At the same time, however, students are also increasingly accused of using AI for cheating~\cite{cotton2023}.
This contradiction, whereby generative AI is used both as an educational resource and an enforcement target, has created a structural conflict that students, teachers, and broader communities are actively negotiating in public forums~\cite{hicke2024,haque2022}.

Large Language Models (LLMs) such as ChatGPT, Claude, and Gemini can scaffold learning, provide feedback, and offer low-friction tutoring~\cite{kasneci2023}, but they can also automate graded work, intensifying institutional anxiety about authorship, cheating, and assessment. The release of ChatGPT in November 2022 triggered rapid adoption in educational contexts, with surveys indicating that over 40\% of college students reported using generative AI for coursework within the first year of availability~\cite{cotton2023}. This adoption has been accompanied by equally rapid deployment of AI-writing detection tools, despite significant concerns about their reliability and fairness~\cite{liang2023,weber2024}.

Naturally, debates around these issues are likely to take place on social platforms and in online communities, making studying these ecosystems both meaningful and useful for understanding them.
In this paper, we focus on Reddit, a social news aggregation and discussion platform where users and stakeholders may routinely share experiences, seek advice, and articulate normative positions about classroom AI practices. 
In particular, we look at education-related subreddits like r/teachers, r/teenagers, and r/highschool, which host extensive discourse in which users provide detailed accounts of lived experience alongside broader arguments about policy and pedagogy~\cite{staudt2018,carpenter2020}.

We investigate how different Reddit communities discuss generative AI in high school education, focusing on learning outcomes, academic integrity and enforcement, AI detection tools, and emotional framing. By explicitly stratifying discourse by community type, we characterize stakeholder-specific patterns and document how weak technical tools (AI detectors) create asymmetric emotional harm under institutional power imbalance. Our analysis draws on systematic data collection from the Reddit comments/submissions archive hosted on Academic Torrents~\cite{academictorrents}, multi-stage filtering for topic relevance validated against human coding ($\kappa = 0.847$), LLM-assisted annotation validated through three-way reliability assessment, and chi-square tests with effect size reporting to quantify group differences.

Prior Reddit analyses of AI discourse typically aggregate sentiment across users, treating discourse as homogeneous~\cite{melton2024,hicke2024}. By contrast, our analysis shows that stakeholder position fundamentally reshapes how AI is framed, regulated, and emotionally experienced. This is not merely a difference in sentiment; rather, it is a governance conflict in which students and teachers occupy structurally different positions. %
Critically, while prior work has highlighted the unreliability of AI detectors~\cite{liang2023,weber2024}, less is known about the emotional cost of detection-based enforcement or how this cost is distributed across stakeholders. We provide the first evidence that detector discourse is associated with elevated negative emotion, with effects significantly larger for students than teachers, documenting not just that detectors may be flawed tools, but also that their deployment creates measurable harm that falls asymmetrically on those with least institutional power.

\myparagraph{Contributions.}
Overall, our findings provide three main contributions to the study of socio-technical governance around AI in education:

\begin{enumerate}
\item \textit{Role-differentiated governance conflict.} We show that stakeholder position (student vs.\ teacher) fundamentally reshapes how AI is framed---not merely in sentiment, but in whether learning trade-offs are articulated at all. Teachers engage in dual framing (simultaneously more optimistic and pessimistic); students discuss AI tactically without explicit learning framing.
\item \textit{Detectors as emotional harm generators.} We document that AI detector discourse is associated with significantly elevated negative emotion (+0.132 risk difference overall, +0.194 for students), providing quantitative evidence that weak enforcement tools create asymmetric harm under institutional power imbalance.
\item \textit{Mechanism-focused case studies.} We include quotes that illustrate \textit{how} quantitative patterns manifest: false accusation escalation, reasoning shifts, and the burden-of-proof asymmetry students face.
\end{enumerate}

\section{Background \& Related Work}

\subsection{Why Study Reddit?}

Reddit is a social news aggregation, content rating, and discussion platform where registered users submit content in the form of text posts, links, images, or videos, which are then voted up or down by other members~\cite{reddit}. 
Content is organized into user-created boards called ``subreddits,'' each dedicated to a specific topic and governed by community-specific rules and volunteer moderators.

Reddit operates with a design around pseudonymity and community self-governance. Users typically maintain persistent usernames but reveal limited personal information, creating a middle ground between full anonymity and real-name policies~\cite{bernstein2011}. This structure has made Reddit a significant venue for discourse on sensitive topics, including education, mental health, and professional practice.~\citet{bernstein2011} characterize this as ``nonymity,'' i.e., a state between anonymity and identified participation that encourages candid discussion while maintaining accountability through reputation.

Research has documented Reddit's role as an ``affinity space'' where users with shared interests congregate to exchange information, experiences, and opinions. %
\citet{gee2017} defines affinity spaces as locations where people interact based on shared activities or interests rather than shared demographic characteristics. Education-related subreddits exemplify this concept: r/teachers functions as a professional community with over 500,000 members where educators discuss pedagogy, policy, and workplace challenges~\cite{staudt2018}; r/teenagers (approximately 3 million members) and r/highschool serve as peer support spaces where students share experiences and seek advice~\cite{carpenter2020}.

\citet{staudt2018} analyzes r/teachers as an affinity space, finding that teachers use the subreddit to share resources, vent frustrations, and seek advice on classroom management and policy issues. 
Also,~\citet{carpenter2020} later document how teachers on Reddit engage in informal professional development through peer interaction. These studies establish Reddit as a legitimate site for studying educator discourse, with the obvious but critical caveat that Reddit users are not demographically representative of the broader teaching profession (as well as of the student population).

\subsection{Related Work}

\myparagraph{Reddit Discourse on Generative AI.} Large-scale analyses of Reddit have begun to map how users discuss AI technologies.~\citet{melton2024} conduct an exploratory study of 5.4 million Reddit submissions using topic modeling and sentiment analysis, identifying a distinct ``Generative AI'' cluster characterized by discussions of image generation, text synthesis, and creative applications. Their work shows how Reddit serves as an early arena for sharing techniques, enthusiasm, and policy discussion around emerging AI technologies.

\citet{hicke2024} examine ChatGPT discussions specifically across 25 education-related subreddits in the months following ChatGPT's release. They characterize stakeholder reactions in terms of volume, engagement, and themes, finding heterogeneous responses that varied by educational role. Teachers express greater concern about academic integrity, while students focus more on practical applications and avoidance of detection. 
\cite{haque2022} analyze Reddit discussions of AI ethics more broadly, finding that users engage with complex ethical questions around bias, fairness, and accountability. Reddit users can engage in nuanced ethical reasoning about AI systems, rather than merely expressing uninformed opinions.

Overall, prior work typically aggregates across user roles and educational levels, leaving open how high school students, teachers, and mixed communities differ in their framing of learning, integrity, and emotional impact when discussing classroom AI. This paper addresses that gap by explicitly stratifying Reddit discourse by community type relevant to high school education.

\myparagraph{Large Language Models in Education.} The integration of LLMs into educational practice has been rapid and contested. 
\citet{kasneci2023} provide a comprehensive overview of opportunities and challenges, identifying potential benefits including personalized tutoring, writing assistance, and accessibility support for students with disabilities. They also catalog risks, including over-reliance on AI, reduced critical thinking, equity concerns for students without access, and the erosion of conventional notions of ``original'' student work.
Then,~\citet{yan2024} present a systematic scoping review of 118 peer-reviewed articles on LLMs in education, identifying practical challenges (integration with existing curricula, teacher training needs, infrastructure requirements) and ethical challenges (privacy, bias, assessment validity, academic integrity). They find that ethical concerns dominate the literature, with academic integrity cited as the most frequently discussed issue.

Empirical studies of student AI use paint a complex picture.~\citet{mollick2023} document how students use ChatGPT for brainstorming, outlining, and revision, often in ways that enhance rather than replace their own thinking. However,~\citet{cotton2023} find that a significant minority of students use AI to generate substantial portions of assessed work without disclosure, creating challenges for assessment validity.

\subsection{Academic Integrity \& Generative AI}
The rise of LLMs has put academic integrity policies under strain. Most existing policies were drafted for plagiarism, i.e., copying from human sources, rather than machine-generated text, creating definitional ambiguity about what constitutes misconduct~\cite{perkins2023}. 
More precisely, LLMs complicate traditional definitions because the technology can produce text that is neither copied nor entirely ``original'' in the conventional sense; misconduct increasingly hinges on disclosure and intent rather than the mere presence of AI-assisted text~\cite{perkins2023}.

Institutional responses have varied widely. Some institutions initially banned ChatGPT entirely, while others integrated it into curricula~\cite{ban2023}.
An analysis of 24 academic integrity policies~\cite{frontiers2024} shows significant inconsistency in how permitted/prohibited AI use is defined, with many policies using vague language that leaves substantial discretion to individual instructors.

\myparagraph{AI-Writing Detectors.} %
Following ChatGPT's release, commercial tools such as Turnitin's AI detection, GPTZero, and Originality.ai have quickly positioned themselves as solutions to the academic integrity challenge. These tools typically use machine to distinguish human-written from AI-generated text based on statistical patterns such as perplexity (how ``surprising'' word choices are) and burstiness (variation in sentence complexity)~\cite{detection2023}.

However, there are significant limitations.
For instance, GPT detectors are systematically biased against non-native English writers: they flagged over 60\% of TOEFL essays written by non-native speakers as AI-generated, compared to less than 10\% of native-speaker essays~\cite{liang2023}, with serious equity implications for diverse student populations. 
Arguably, this bias stems from non-native writers producing text with lower perplexity (more predictable word choices) due to limited vocabulary, which detectors interpret as a signal of AI-generated text.

~\citet{weber2024} present a systematic evaluation of 14 AI-writing detectors across multiple academic contexts, finding accuracy rates that varied widely (from 55\% to 85\%) and false positive rates ranging from 5\% to over 30\% depending on the tool and text type. 
Similarly,~\citet{elkhatat2023} find that AI detectors perform poorly on mixed human-AI text (where students use AI for some portions and write others themselves), which is likely the most common real-world use case, while~\citet{sadasivan2023} demonstrate that simple paraphrasing techniques can evade detection with high success rates, undermining the deterrent value of detection-based approaches.

Institutional guidance has increasingly cautioned against reliance on detectors, with notable examples coming from ~\citet{iowa2024},~\citet{niu2024}, and~\cite{sandiego2024}, which summarizes the evidence as showing detectors are ``neither accurate nor reliable'' for high-stakes decisions.

\begin{figure}[t]
\centering
\includegraphics[width=0.99\columnwidth]{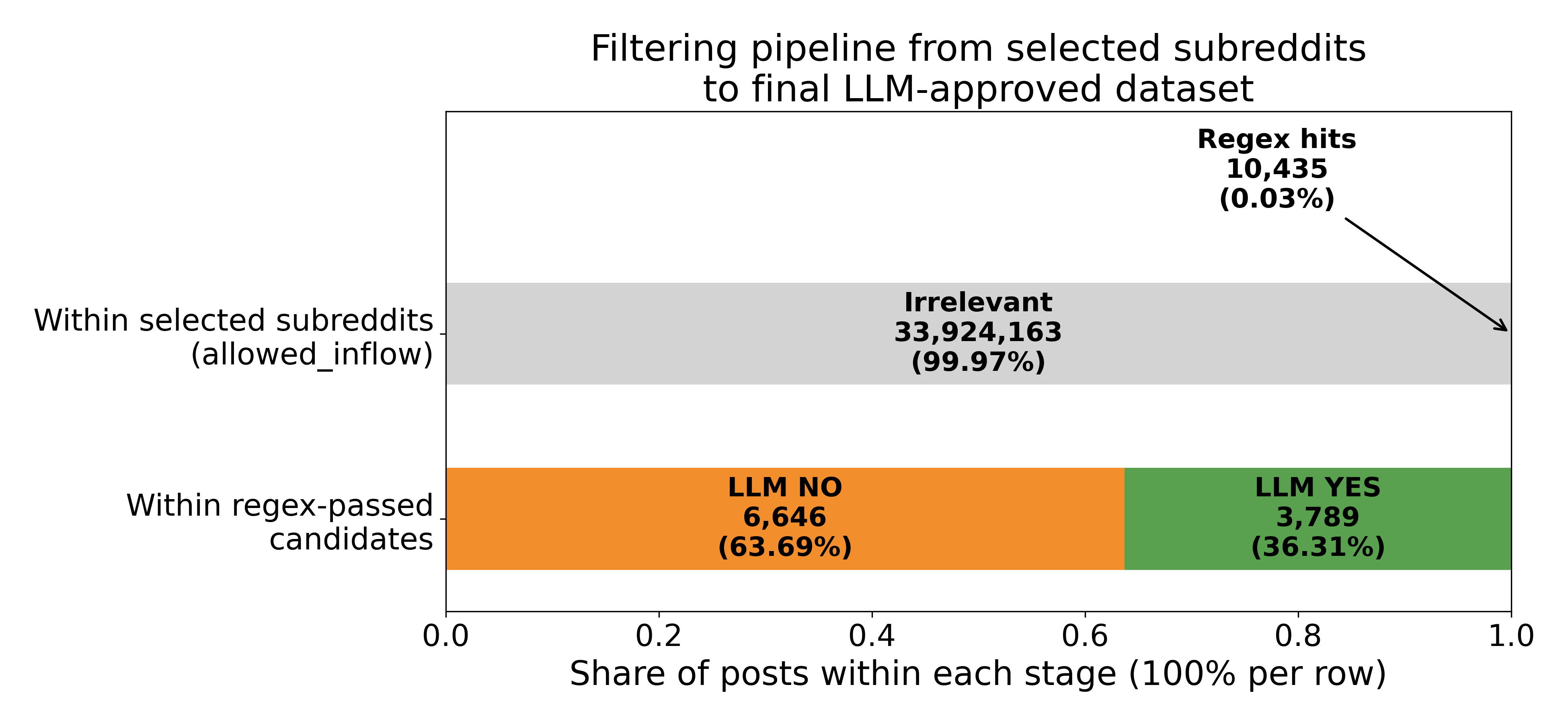}
\caption{Data filtering pipeline. From 33.9M posts in selected subreddits, keyword matching identified 10,435 candidates, of which 3,789 (36.31\%) were classified as relevant to generative AI in education by the LLM filter.}
\label{fig:pipeline}
\end{figure}

\section{Data and Methods}\label{sec:data}
In this section, we describe our data collection, filtering, annotation, and statistical analysis procedures.

\subsection{Dataset}\label{sec:dataset}

We retrieve posts from the Reddit comments/submissions archive hosted on Academic Torrents~\cite{academictorrents} spanning two years---more precisely, from December 2022 to December 2024. This 24-month period captures the initial diffusion of ChatGPT and similar LLMs into educational settings following ChatGPT's public release in November 2022, as well as the subsequent deployment of AI detection tools and evolution of institutional responses across two full academic years.

Queries combined subreddit filters with generative AI keywords: \texttt{ChatGPT}, \texttt{GPT}, \texttt{AI essay}, \texttt{AI writing}, \texttt{Turnitin AI}, \texttt{AI detector}, \texttt{AI homework}, and \texttt{AI cheating}. Education markers included: \texttt{homework}, \texttt{essay}, \texttt{assignment}, \texttt{teacher}, \texttt{high school}, \texttt{class}, and \texttt{grade}. Inclusion criteria required that (i) the text referenced generative AI systems in an educational context, and (ii) the post or comment was in English.

Figure~\ref{fig:pipeline} illustrates the filtering pipeline. From 33.9 million posts in selected subreddits, keyword regex matching identified 10,435 candidate posts (0.03\%). These candidates were then filtered for topic relevance using an LLM classifier (see Section~\ref{sec:topic_filtering}), yielding a final dataset of 3,789 posts (36.31\% of regex-matched candidates).
More precisely:

\begin{itemize}
\item Student communities (r/teenagers, r/highschool): 204 (5.38\%)
\item Teacher communities (r/teachers): 2,769 (73.08\%)
\item Mixed communities (r/chatgpt, r/askteachers): 816 (21.54\%)
\end{itemize}

\begin{figure}[t]
\centerline{\includegraphics[width=0.9\columnwidth]{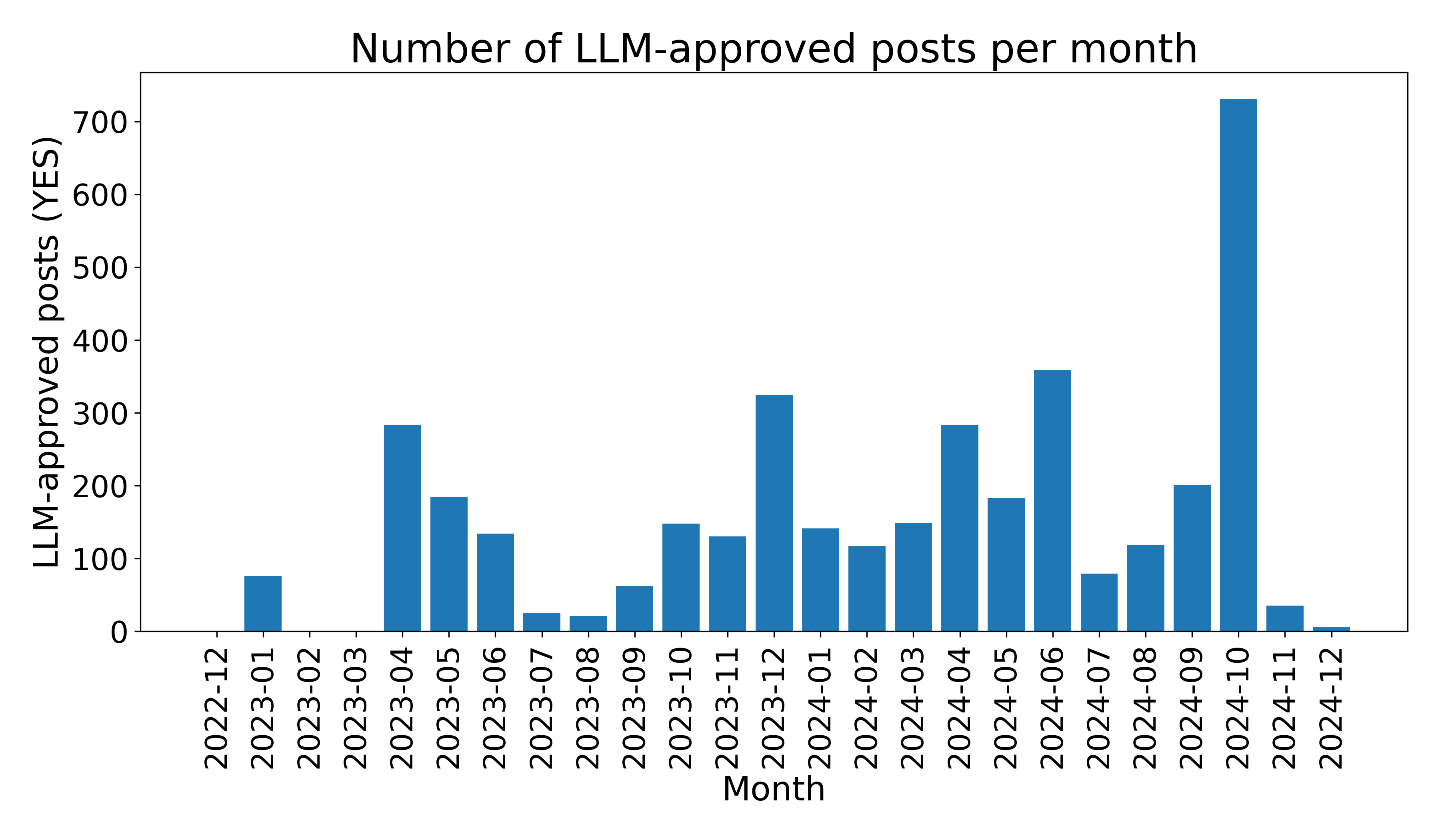}}
\caption{Monthly distribution of relevant posts (January 2023 to October 2024).}
\label{fig:temporal}
\end{figure}

\myparagraph{Temporal distribution.} Figure~\ref{fig:temporal} shows the monthly distribution of posts. Discourse volume exhibits clear academic-year seasonality: posts decline sharply during summer months (July to August 2023 and 2024) and peak during the school year, particularly in spring semesters (April 2023, April 2024) and fall back-to-school periods (October 2024). This pattern validates that the discourse is tied to school activity cycles rather than general Reddit trends, and suggests that AI-in-education concerns are most salient when students and teachers are actively engaged in classroom work.

\myparagraph{Implication of group imbalance.} Teachers constitute 73\% of the dataset, reflecting differences in posting volume across subreddits rather than sampling bias: r/teachers is a substantially larger and more active community than student-focused subreddits. This imbalance means that (i) student estimates have wider uncertainty due to smaller samples, (ii) overall patterns are dominated by teacher discourse, and (iii) effects may be attenuated if teacher-student differences exist but teacher posts overwhelm student signal. We address this by emphasizing within-group comparisons and effect sizes rather than raw prevalence.

\subsection{Topic Relevance Filtering}\label{sec:topic_filtering}
Our keyword-based retrieval produces many off-topic posts (e.g., general ChatGPT discussions without educational context, mentions of AI in non-academic settings). To filter for relevance, we code each candidate post as either substantively discussing generative AI in an educational context or not.
A human coder (first author) and an LLM (GPT-4) independently coded a random sample of 150 posts for topic relevance (binary: relevant/not relevant).

\begin{table}[t]
\centering
\small
\begin{tabular}{lr}
\toprule
\bf Metric & \bf Value \\
\midrule
Cohen's $\kappa$ & 0.847 \\
Percent agreement & 93.3\% \\
Human prevalence (relevant) & 68.0\% \\
LLM prevalence (relevant) & 66.7\% \\
\bottomrule
\end{tabular}
\caption{Topic relevance filtering: Human-LLM agreement on 150-post sample.}
\label{tab:topic_reliability}

\end{table}

In Table~\ref{tab:topic_reliability}, we report agreement.
Given high agreement ($\kappa = 0.847$, indicating almost perfect agreement under standard benchmarks~\cite{landis1977}), we use the LLM to filter the full candidate set. 
We exclude posts classified as not relevant, yielding the final dataset of 3,789 posts.

\subsection{Validity of Subreddit-as-Stakeholder Proxy}\label{sec:proxy}

A core methodological assumption is that subreddit membership serves as a reasonable proxy for stakeholder identity, i.e., that users posting in r/teachers are predominantly teachers, and users posting in r/teenagers or r/highschool are predominantly students. 
We validate this assumption through manual review of stratified random samples.

\myparagraph{Teacher validation ($n=100$).} To validate the teacher set, one coder has independently reviewed 100 randomly sampled posts from r/teachers, coding for explicit self-identification as a teacher, educator, or school staff member. Markers include: first-person references to ``my students,'' ``my classroom,'' or ``my school''; statements like ``As a [subject] teacher...''; discussion of grading, lesson planning, or administrative duties from an educator perspective; and use of teacher-specific user flair. 
Ultimately, we find that 94\% of posts contain explicit educator self-identification.

\myparagraph{Student validation ($n=50$).} Next, a coder reviews 50 randomly sampled posts from r/teenagers and r/highschool, coding for student self-identification. Markers include references to ``my teacher,'' ``my class,'' or ``my homework''; discussion of grades, assignments, or school experiences from a student perspective; age-related context (e.g., ``I'm a junior,'' ``I'm 16''). 
Similarly, 88\% of posts contain explicit student self-identification.

\myparagraph{Limitations.} We acknowledge that misclassification may naturally occur, as some teachers may post in student subreddits seeking student perspectives, and students likely lurk and occasionally post in r/teachers. This misclassification would \textit{attenuate} observed group differences (tending to bias toward the null), meaning true stakeholder differences may be larger than reported.
However, we believe this effect to be minimal.

\subsection{Label Framework}\label{framework}
We then annotate each post with a non-mutually exclusive set of binary labels indicating the presence or absence of specific frames. 
We use labels developed through iterative coding of an initial sample, drawing on prior work on AI discourse~\cite{melton2024,hicke2024} and educational technology governance~\cite{kasneci2023,yan2024}.
More precisely, labels fall into three families, as discussed below.

\myparagraph{Learning-Impact Labels:}
\begin{itemize}
\item helps\_learning: Claims/implies AI improves learning, understanding, or educational outcomes.
\item harms\_learning: Claims/implies AI undermines learning, creates dependency, or degrades educational quality.
\item tutoring\_learning\_support: Describes AI as a tutoring tool, feedback mechanism, or learning scaffold.
\item Potential\_to\_help: Speculates AI could help learning under certain conditions or with proper implementation.
\item positive\_learning\_effect: Reports concrete positive educational outcomes from AI use.
\end{itemize}

\myparagraph{Integrity and Enforcement Labels:}
\begin{itemize}
\item cheating\_academic\_integrity: Discusses cheating, academic integrity policies, or misconduct concerns.
\item AI\_Detector\_Related: References AI-writing detection tools, detection scores, or detection-based accusations.
\end{itemize}

\myparagraph{Emotional Framing Labels:}
\begin{itemize}
\item fear\_anxiety: Expresses worry, anxiety, or apprehension about AI in education.
\item anger\_frustration: Expresses anger, frustration, or resentment about AI-related situations.
\item excitement\_hope: Expresses enthusiasm, optimism, or hope about AI in education.
\item negative\_emotion: Indicates presence of fear, anxiety, anger, frustration, or other negative affect.
\end{itemize}

\subsection{Annotation Procedure and Reliability}
Our next step is to annotate each post with the label framework described in Section~\ref{framework}, which captures learning impact, integrity and enforcement, and emotional framing. We do this in two stages: 1) establishing human baseline reliability, and 2) validating LLM annotation for scalable coding.

\myparagraph{Three-Way Reliability Assessment.}
To validate the annotation approach, three coders independently annotated a stratified random sample of 100 posts: the first author (Coder A), a senior author (Coder B), and an LLM (GPT-4) using the annotation guide as a system prompt. This three-way comparison allows assessment of both human-human and LLM-human agreement.
In Table~\ref{tab:reliability_comparison}, we report the corresponding agreement scores.

\begin{table}[t]
\centering
\small
\begin{tabular}{lrrr}
\toprule
\textbf{Label} & \textbf{A/LLM} & \textbf{B/LLM} & \textbf{A/B} %
\\
\midrule
AI\_Detector\_Related         & 0.741 & 0.810 & --- %
\\
positive\_learning\_effect    & 0.697 & 0.382 & 0.422 %
\\
tutoring\_learning\_support   & 0.680 & 0.382 & 0.287 %
\\
Potential\_to\_help           & 0.667 & 0.269 & 0.242 %
\\
helps\_learning               & 0.643 & 0.554 & 0.512 %
\\
fear\_anxiety                 & 0.639 & 0.392 & 0.419 %
\\
cheating\_academic\_integrity & 0.603 & 0.618 & 0.580 %
\\
negative\_emotion             & 0.600 & 0.541 & 0.460 %
\\
anger\_frustration            & 0.513 & 0.375 & 0.358 %
\\
harms\_learning               & 0.470 & 0.351 & 0.415 %
\\
excitement\_hope              & 0.459 & 0.558 & 0.473 %
\\
\midrule
\textbf{Macro Average}        & \textbf{0.610} & \textbf{0.476} & \textbf{0.417} %
\\
\bottomrule
\end{tabular}
\caption{Agreement (Cohen's $\kappa$) on the 100-post overlap set: Coder A vs.~LLM, Coder B vs.~LLM, and A vs.~B.}
\label{tab:reliability_comparison}
\end{table}

The three-way comparison reveals an important pattern: LLM-human agreement (macro $\kappa = 0.48$--$0.61$) is comparable to or higher than human-human agreement (macro $\kappa = 0.42$). This suggests that disagreement reflects genuine ambiguity in the annotation task rather than LLM-specific error. The LLM shows particularly strong agreement on key labels central to our research questions: AI\_Detector\_Related ($\kappa = 0.74$--$0.81$) and cheating\_academic\_integrity ($\kappa = 0.58$--$0.62$).
Human-human agreement of $\kappa = 0.42$ falls in the moderate range under standard benchmarks~\cite{landis1977}, reflecting the inherent difficulty of coding nuanced frames like Potential\_to\_help ($\kappa = 0.24$) from naturalistic social media text. This level of agreement is consistent with prior content analysis work on subjective constructs in online discourse~\cite{melton2024,haque2022}.

\myparagraph{Annotation Decision.} Given that (i) LLM-human agreement equals or exceeds human-human agreement, (ii) the LLM shows strong agreement on high-stakes labels central to our research questions, and (iii) manual annotation of 3,789 posts was infeasible, we code the full dataset using the LLM with the same annotation guide.

Arguably, replacing one human annotator with an LLM does not reduce reliability relative to standard qualitative practice when LLM-human $\kappa$ meets or exceeds human-human $\kappa$. The LLM functions as a measurement instrument for scaling a human-validated coding scheme, not as an interpreter generating novel categories.
We emphasize that our key claims---the detector$\rightarrow$emotion association and the student-teacher difference in detector discussion rates---rely exclusively on high-agreement labels. AI\_Detector\_Related shows the strongest agreement across all coder pairs ($\kappa = 0.74$--$0.81$), and negative\_emotion shows moderate-to-substantial agreement ($\kappa = 0.46$--$0.60$). Claims based on lower-agreement labels (e.g., Potential\_to\_help, $\kappa = 0.24$--$0.27$) are treated as suggestive rather than definitive.

\myparagraph{Remarks.} 
Nevertheless, we acknowledge that all coders (human and LLM) show lower agreement on nuanced labels. We interpret results for these labels cautiously and pair quantitative patterns with qualitative exemplars. The directional validity of our findings is supported by convergent evidence: statistical associations, qualitative mechanisms, and alignment with prior literature on detector harms~\cite{liang2023,weber2024,niu2024}.

\subsection{Statistical Analysis}

For each label, we test whether prevalence differs across author groups using a $3\times2$ chi-square test of independence. We report chi-square statistics, Cram\'er's $V$ effect sizes, and Benjamini-Hochberg adjusted $q$-values to control false discovery rate across multiple comparisons.

\myparagraph{Independence assumption and robustness:} Chi-square assumes independent observations. Reddit users can post multiple times, and threaded discussions create dependence. We address this limitation in three ways:

\begin{enumerate}
\item We treat results as \textit{descriptive inference about discourse patterns} rather than individual-level causal claims. The unit of analysis is the post/comment, not the user.
\item For the primary detector$\rightarrow$emotion association, we rerun the analysis after deduplicating by author (retaining each author's first post). The association remained significant ($\chi^2 = 38.2$, $p < 10^{-9}$, $\phi = 0.11$), with risk difference attenuated from +0.132 to +0.108.
\item We emphasize effect sizes (Cram\'er's $V$, $\phi$, risk differences) over $p$-values, since effect sizes are less sensitive to sample size inflation from repeated authors.
\end{enumerate}

For detector-emotion associations, we construct $2\times2$ contingency tables and compute $\phi$ coefficients and risk differences as follows:
\begin{align*}
\small
\text{Risk difference} & = P(\text{neg} = 1 \mid \text{detector} = 1) +\\ 
& - P(\text{neg} = 1 \mid \text{detector} = 0)
\end{align*}

\section{Results}\label{sec:results}
In this section, we report our experimental evaluation.
More precisely, we quantify overall label prevalence, group differences across stakeholder communities, the association between detector-related discourse and negative emotion, as well as differences in learning-stance distributions.

\begin{figure}[t]
\centerline{\includegraphics[width=\columnwidth]{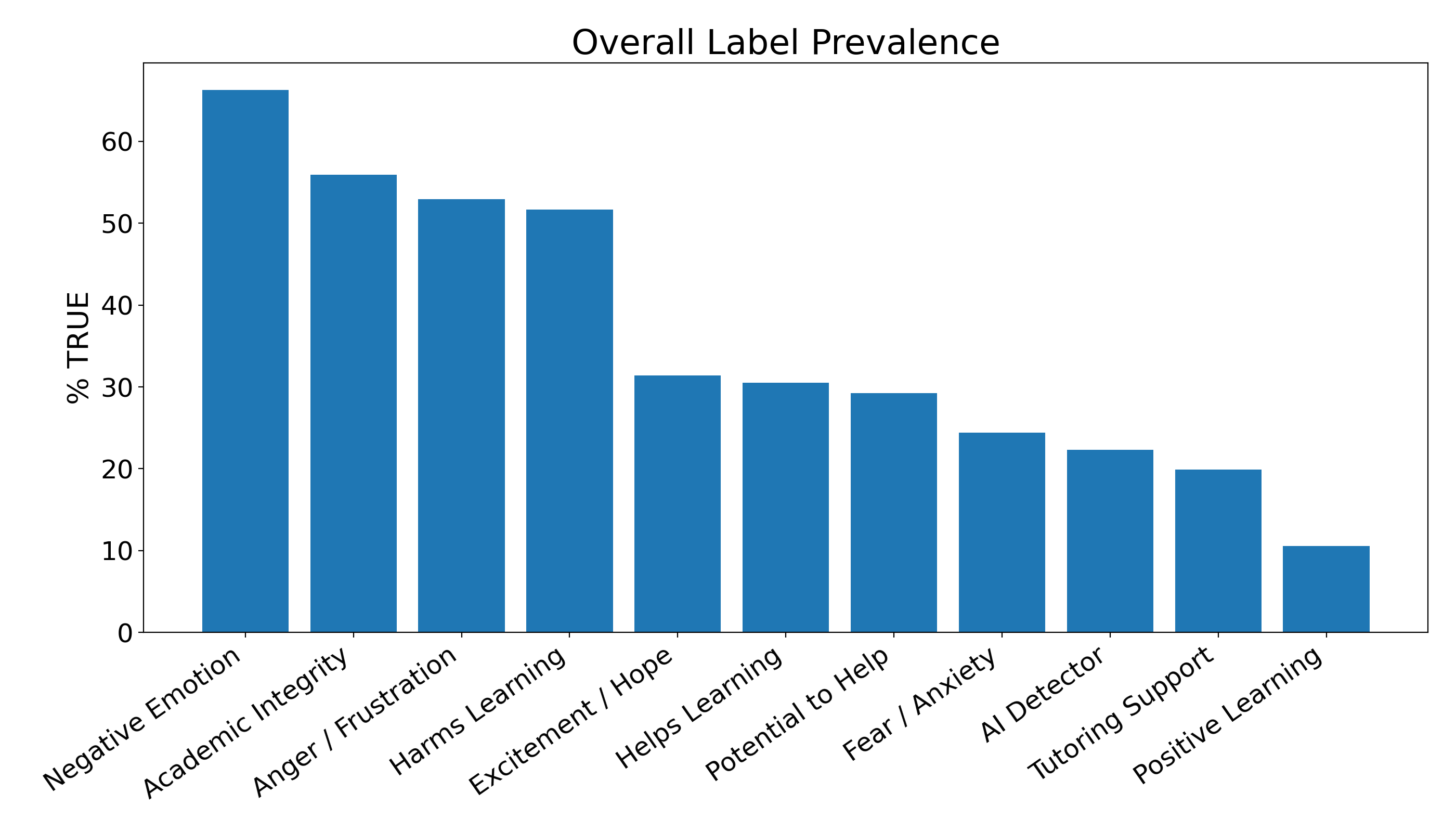}}
\caption{Overall label prevalence across all 3,789 posts. Discourse is dominated by negative emotions and concerns about academic integrity, while positive learning outcomes are rare. }
\label{fig:prevalence}
\end{figure}

\subsection{Overall Label Prevalence}
We begin by examining overall label prevalence across all posts.
Across all 3,789 posts, discourse is dominated by negative emotion (66.30\%) and integrity concerns (55.90\%), while explicit positive learning outcomes are rare (10.53\%). 
In Figure~\ref{fig:prevalence}, we report the full distribution of labels.

This distribution indicates that Reddit discourse about generative AI in education is characterized more by concern, conflict, and negative affect than by enthusiasm or reported benefits.
This is consistent with the contested nature of AI in educational governance~\cite{kasneci2023,yan2024}.

\subsection{Group Differences Across Labels}

Next, in Table~\ref{tab:chisquare}, we report the chi-square tests for group differences. Several learning and enforcement frames show statistically significant differences with modest effect sizes (Cram\'er's $V$ in the 0.07--0.12 range).

\begin{table}[t]
\centering
\small
\setlength{\tabcolsep}{4pt}

\begin{tabular}{lrrrrr}
\toprule
Label & Stu\% & Tea\% & Mix\% & $V$ & $q$ \\
\midrule
harms\_learning & 39.7 & 55.2 & 42.8 & 0.116 & $<10^{-10}$ \\
Potential\_to\_help & 10.8 & 31.7 & 25.7 & 0.111 & $<10^{-9}$ \\
helps\_learning & 12.8 & 32.2 & 29.2 & 0.096 & $<10^{-7}$ \\
cheating\_acad\_int & 55.4 & 58.2 & 48.3 & 0.081 & $<10^{-5}$ \\
AI\_Detector\_Rel & 31.9 & 20.4 & 26.5 & 0.081 & $<10^{-5}$ \\
excitement\_hope & 17.7 & 32.1 & 32.4 & 0.071 & $1.4\times10^{-4}$ \\
\bottomrule
\end{tabular}
\caption{Group prevalence by label with chi-square tests and BH correction.}
\label{tab:chisquare}
\end{table}

\myparagraph{Teacher Dual Framing.} We find that teachers show higher rates of harms\_learning (55.2\%) than students (39.7\%), with $\chi^2(2) = 51.26$, $p < 10^{-11}$, $V = 0.116$. However, teachers \textit{also} show higher rates of helps\_learning (32.2\% vs.\ 12.8\%) and Potential\_to\_help (31.7\% vs.\ 10.8\%). This dual framing -- i.e., simultaneously more pessimistic and more optimistic -- suggests teachers are compelled by institutional responsibility to articulate explicit positions on AI's learning impact, both positive and negative. Students, by contrast, often discuss AI without explicit learning framing (49.5\% in ``neither helps nor harms'' bucket; see Section 4.4).

\begin{figure}[t]
\centerline{\includegraphics[width=0.7\columnwidth]{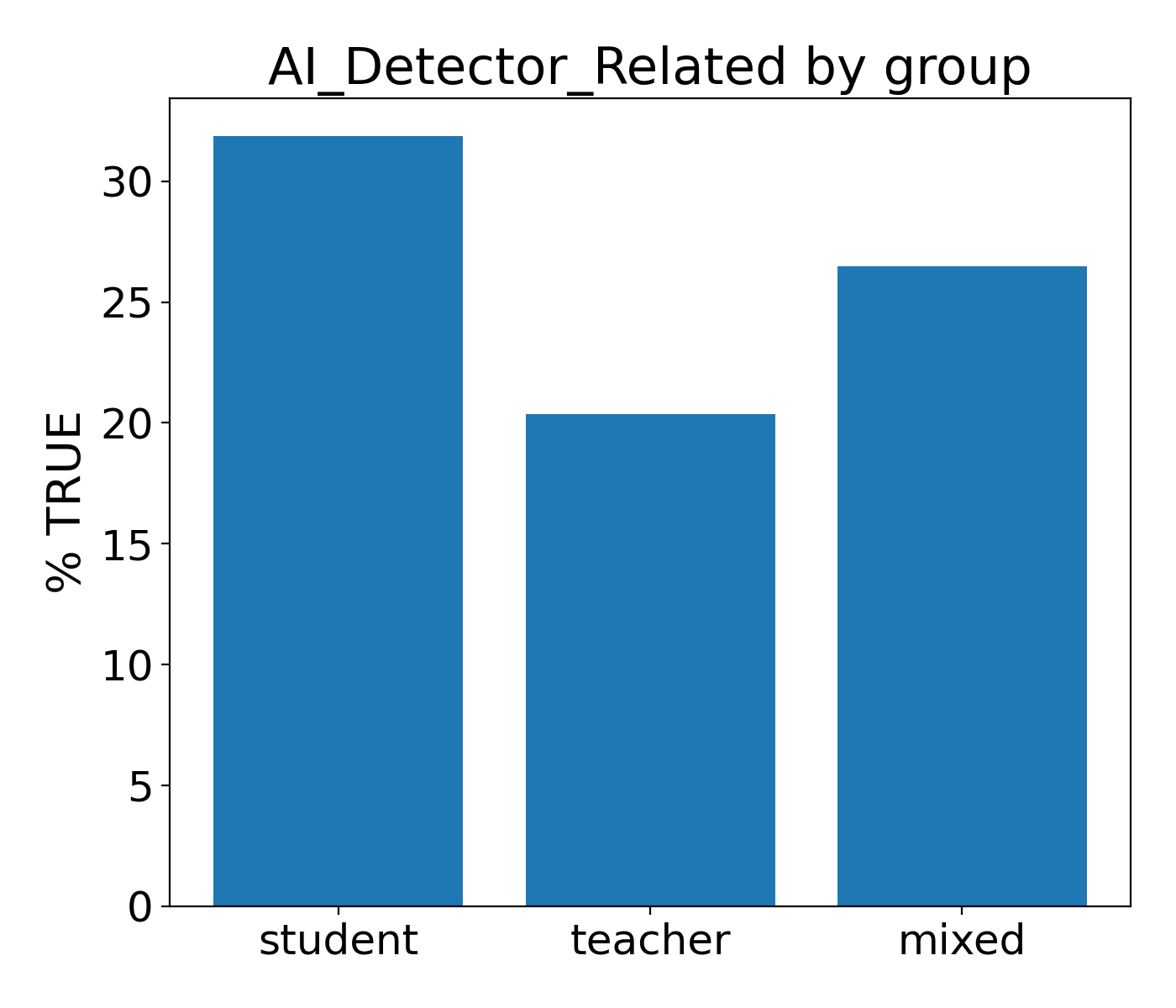}}
\caption{AI detector discussion by stakeholder group. Students discuss detectors at higher rates than teachers, consistent with students being more directly subject to detection-based enforcement.}
\label{fig:detector_group}
\end{figure}

\myparagraph{Discourse around Detectors.} Students discuss AI detectors at higher rates (31.9\%) than teachers (20.4\%), with $\chi^2(2) = 24.92$, $p < 10^{-5}$, consistent with students being more directly affected by detector-based enforcement (Figure~\ref{fig:detector_group}). This finding is supported by the label with the highest LLM-human agreement ($\kappa = 0.74$--$0.81$), lending confidence to the inference. However, given the smaller student sample ($n = 204$), we treat student-specific prevalence estimates as suggestive of directional differences rather than precise population parameters.

\begin{table}[t]
\centering
\small
\begin{tabular}{lrrrr}
\toprule
\bf Scope & $n$ & \bf Risk Diff. & $\chi^2$ & $\phi$ \\
\midrule
Overall & 3,789 & +0.132 & 51.34 & 0.116 \\
Student & 204 & +0.194 & 7.19 & 0.188 \\
Teacher & 2,769 & +0.103 & 21.05 & 0.087 \\
Mixed & 816 & +0.200 & 29.37 & 0.190 \\
\midrule
\textit{Deduplicated} & 2,847 & +0.108 & 38.2 & 0.110 \\
\bottomrule
\end{tabular}
\caption{Association between detector talk and negative emotion.}
\label{tab:detector}
\end{table}

\subsection{Detector Talk and Negative Emotion}
We then measure the association between detector-related discussion and negative emotion; see Table~\ref{tab:detector}.
We find that detector-related posts are significantly more likely to include negative emotion than non-detector posts ($\chi^2 = 51.34$, $p < 10^{-12}$). The probability of negative emotion given detector talk is 0.766 versus 0.634 for non-detector posts (risk difference +0.132). Risk differences are larger for students (+0.194) and mixed communities (+0.200) than for teachers (+0.103), suggesting that detection discourse may be a particularly strong trigger for negative emotion among those subject to enforcement rather than those administering it. Given the smaller student sample, we emphasize the consistent \textit{direction} of this effect across all groups rather than precise magnitude estimates for students specifically.

The association remains significant after author deduplication (bottom row), with risk difference attenuated from +0.132 to +0.108. This suggests the finding is not solely driven by prolific users posting multiple detector-related complaints.
Note that both AI\_Detector\_Related ($\kappa = 0.74$--$0.81$) and negative\_emotion ($\kappa = 0.54$--$0.60$) show moderate-to-strong LLM-human agreement, supporting confidence in this finding.

\subsection{Learning-Stance Distribution}
\begin{figure}[t]
\centerline{\includegraphics[width=0.85\columnwidth]{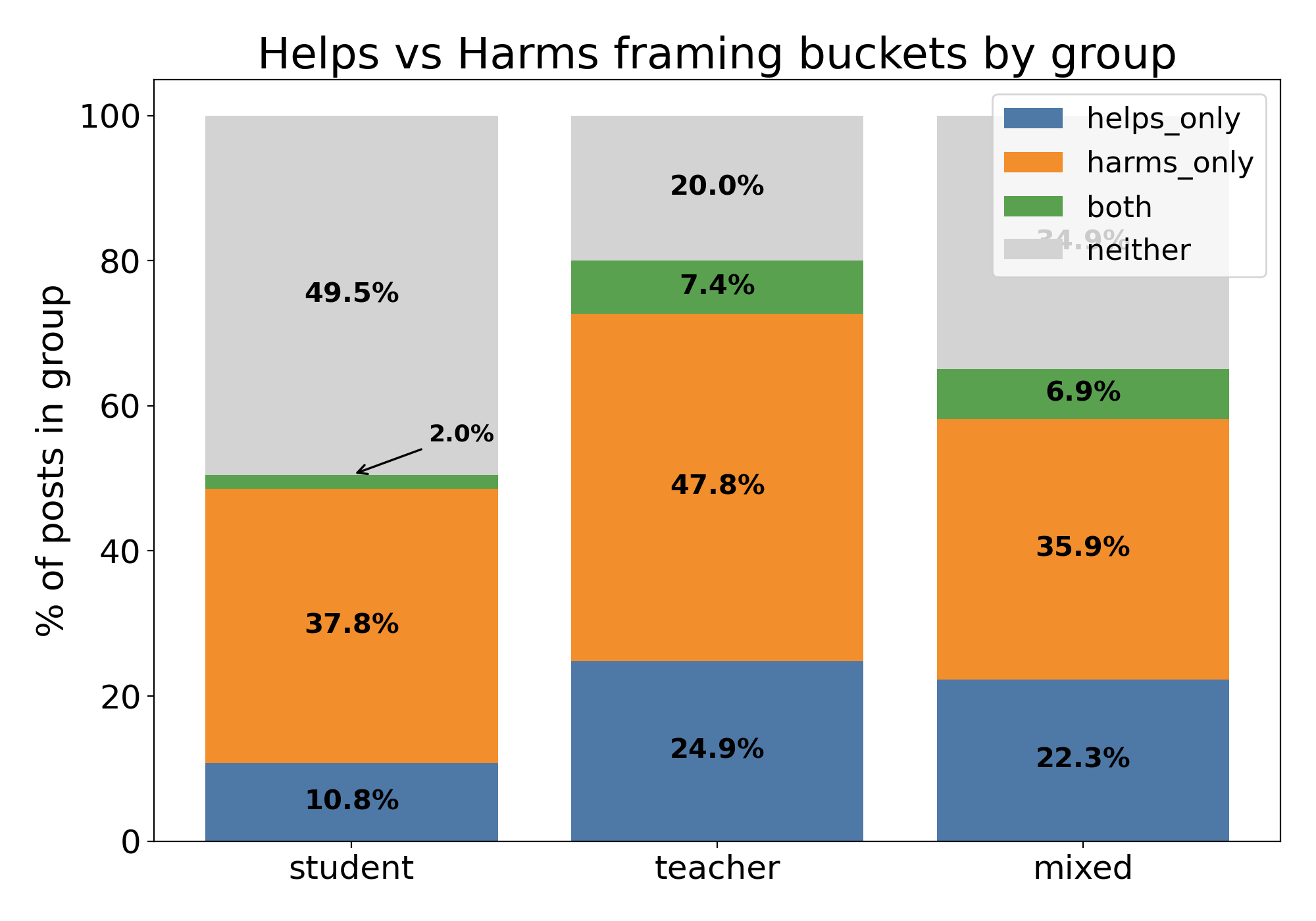}}
\caption{Learning stance distribution by stakeholder group. Students show the largest ``neither'' share (red), discussing AI without explicit learning framing. Teachers more frequently articulate explicit trade-offs (``both,'' green) and positions on AI's learning impact.}
\label{fig:buckets}
\end{figure}

Finally, we analyze the distribution of learning stances across stakeholder groups. Figure~\ref{fig:buckets} visualizes this distribution.

Group-by-bucket differences are statistically significant ($\chi^2(6) = 157.03$, $p < 10^{-30}$, Cramer's $V = 0.144$). Teachers are more likely to articulate explicit trade-offs (7.4\% ``both helps and harms'') compared to students (2.0\%), while students have the largest ``neither'' share (49.5\%), discussing AI without explicit learning-impact framing. This pattern is consistent with the interpretation that teachers engage AI discourse through a pedagogical lens shaped by institutional responsibility, whereas students engage it through an enforcement-and-consequence lens shaped by their position as subjects of institutional power.

\section{Case Studies}

To illustrate the mechanisms behind quantitative patterns, we now present and discuss a few case studies, reporting (anonymized) quotes that explain \textit{how} the statistical associations manifest in lived experience.

\subsection{Student Discourse: False Accusations and Detector Anxiety}

Students discuss AI detectors at significantly higher rates than teachers (31.9\% vs.\ 20.4\%), and detector-related posts show the largest negative emotion elevation among students (+0.194 risk difference). The following excerpts illustrate the mechanisms driving these patterns.

A common mechanism in our dataset relates to students identifying or discussing (possibly latent) institutional failures.
For instance, teachers may often misunderstand LLM behavior and default to escalating the issue to administrative involvement, ultimately placing the burden on the student to explain the technical limitations.
This pattern is exemplified by the following quote:

\begin{quote}
\textit{``my teacher straight up just asked chatgpt, `did you write [username]'s essay?' and the AI replied `yes I generated this text'... I had to have video meeting with multiple teachers and school board members to explain how ai and ai detectors work.''}   \textbf{\em -- Student, r/teenagers}
\end{quote}

As students bear asymmetric consequences for institutional misunderstanding, negative emotion is likely to emerge in detector-related discourse.
Moreover, students often express ``anticipatory'' anxiety, i.e., fear about future false positives rather than a response to an actual incident. 
In other words, detector deployment creates ambient stress even for students who have not been accused, contributing to elevated negative emotion, aligning with concerns raised by NIU's teaching center about the ``chilling effect'' of unreliable detection systems~\cite{niu2024}.
For instance:

\begin{quote}
\textit{``My teachers brought that up and that concerned me because idk if the Ai detectors will be false and this year will just be filled with awful grades because of it.''}   \textbf{\em -- Student, r/teenagers}
\end{quote}

\begin{quote}
\textit{``an assignment I turned in... was flagged with 96\% AI on Turnitin... I didn't copy paste anything and I'm scared this will happen to me in the future.''}   \textbf{\em -- Student, r/highschool}
\end{quote}

The latter post provides a concrete case of a high-confidence false positive, consistent with detector reliability limitations documented in peer-reviewed literature~\cite{liang2023,weber2024}. The 96\% score on originality indicates that detector outputs can be both confident and wrong, which creates lasting anxiety about future assignments.

A second mechanism involves enforcement asymmetry, where students who actually use AI may go undetected while students writing original work face false accusations. This undermines both the deterrent effect and perceived fairness of detection systems, as illustrated by the following post:

\begin{quote}
\textit{``I've literally used ChatGPT for the last like 3 essays I've written. None got a bad grade nor did the teacher suspect that it was AI-written.''}   \textbf{\em -- Student, r/teenagers}
\end{quote}

\subsection{Teacher Discourse: Dual Framing and Evidentiary Reasoning}

On the other hand, teachers show elevated rates of \textit{both} harms-learning (55.2\%) and helps-learning (32.2\%) discourse compared to students (39.7\% and 12.8\%), indicating dual framing. 
In our dataset, we observe several examples of how teachers articulate trade-offs and develop alternative enforcement strategies.
Some report pragmatic adaptation, acknowledging that enforcement is impractical, combined with pedagogical reorientation. 
This may (at least in part) explain \textit{why} teachers show dual framing: institutional responsibility compels explicit engagement with both risks and opportunities, leading to the ``tool not crutch'' framing recommended by scholars like Sullivan et al.~\cite{sullivan2023}, and is quite clearly exemplified by the following post:

\begin{quote}
\textit{``It's a losing battle. They are going to use it, and I can't stop that. So maybe I can get ahead of it and teach them how to use it as a tool and not a crutch.''}   \textbf{\em -- Teacher, r/teachers}
\end{quote}

We also observe a varying degree of sophisticated evidentiary reasoning, with reframing from detecting AI to verifying authorship via process artifacts (drafts, revision history, in-class samples) and acknowledging detector unreliability while maintaining enforcement capacity. This shift from product-based to process-based evidence aligns with assessment redesign recommendations~\cite{sullivan2023,redesign2023}.
For example:

\begin{quote}
\textit{``I don't have to prove that an essay was AI-generated; I just have to prove the student didn't write it.''}   \textbf{\em -- Teacher, r/teachers}
\end{quote}

Additional posts provide some anecdotal confirmation of our finding that even teachers, who have lower detector-emotion association (+0.103), express skepticism about detection-based approaches, consistent with institutional guidance cautioning against detector reliance~\cite{turnitin2024,iowa2024,niu2024}:

\begin{quote}
\textit{``AI detectors should never be used. They are like dowsing rods... If your writing is in any way academic it will be flagged as AI.''}   \textbf{\em -- Teacher, r/teachers}
\end{quote}

The ``dowsing rod'' metaphor frames detectors as fundamentally invalid rather than merely imperfect. 

Finally, several posts reveal teachers' own use of AI for administrative tasks, illustrating the potential for an asymmetrical tension between policies restricting student AI use and educators' reliance on AI for professional efficiency. %
For instance:

\begin{quote}
\textit{``I couldn't be arsed to formulate a reply so I threw this prompt into chatGPT... and sent the email.''}   \textbf{\em -- Teacher, r/teachers}
\end{quote}

\subsection{Mixed Community Discourse}

A central mechanism in mixed communities is the burden-of-proof asymmetry inherent in detection-based enforcement, in which students may be required to prove innocence without clear evidentiary standards.

Some posts illustrate the burden-of-proof asymmetry central to student negative emotion: even process evidence may be dismissed, leaving students with no clear path to exoneration.
For example:

\begin{quote}
\textit{``I have shown my version history on google docs... but he still does not believe me... How can I prove my innocence?''}   \textbf{\em -- Mixed, r/chatgpt}
\end{quote}

The question ``How can I prove my innocence?'' captures the Kafkaesque quality of detection-based enforcement when the accused lacks credible means of defense.

\begin{quote}
\textit{``They aren't actually a `scam', but they are incredibly unreliable, and have a high rate of false positives.''}   \textbf{\em -- Mixed, r/chatgpt}
\end{quote}

This post offers nuanced detector assessment, distinguishing ``scam'' (fraudulent intent) from ``unreliable'' (technical limitation)---reflecting sophisticated engagement consistent with the moderate negative emotion elevation in mixed communities (+0.200 risk difference).

\section{Discussion}\label{sec:discussion}

\subsection{Teacher Dual Framing as Key Differentiator}

A central finding is that teachers show elevated rates of \textit{both} harms-learning and helps-learning discourse compared to students. This dual framing %
distinguishes teacher discourse from student discourse and suggests that, under institutional responsibility, teachers are compelled to articulate explicit positions on AI's educational impact.

This finding differs from prior work that characterizes stakeholder reactions as simply ``positive'' or ``negative''~\cite{melton2024,hicke2024}. Teachers appear to engage in genuine weighing of trade-offs, consistent with their professional obligation to consider pedagogical implications. Students, by contrast, more often discuss AI in tactical terms (accusations, grades, detection avoidance) without explicit learning framing —- reflecting their position as subjects of institutional enforcement rather than agents of pedagogical decision-making.

As a result, the implication for policy is that teacher and student concerns may be more compatible than they appear: teachers worry about learning outcomes, students worry about fair treatment, and both could be addressed by assessment redesign approaches that reduce reliance on detection~\cite{sullivan2023,redesign2023}.

\subsection{Detectors as Governance Trigger for Emotional Harm}

The consistent association between detector discourse and negative emotion is, arguably, our biggest finding. Combined with peer-reviewed evidence of detector unreliability~\cite{liang2023,weber2024}, this suggests current detection practices create emotional harm that:

\begin{enumerate}
\item Falls unevenly across stakeholder groups (larger effects for students and mixed communities)
\item Occurs regardless of actual AI use (false positives generate the same emotional response as accurate detection)
\item May undermine educational relationships and trust between students and teachers
\end{enumerate}

Our case studies shed light on the mechanisms: students describe false accusations, the burden of proving innocence, and anticipatory anxiety about future assignments. Even teachers who are skeptical of detectors acknowledge that detector outputs create adversarial dynamics that interfere with educational relationships.

This finding aligns with and extends institutional guidance urging caution in detector deployment~\cite{turnitin2024,iowa2024,niu2024}. Our work documents the emotional dimension of this harm at scale, showing that detector discourse is not merely contested but emotionally charged, suggesting real psychological costs.

\subsection{Stakeholder Misalignment and Policy Implications}

Teachers seem to frame AI policies around learning outcomes; students experience those policies primarily as enforcement mechanisms with high-stakes personal consequences. This misalignment suggests the need for explicit communication that addresses both pedagogical rationale (why AI policies exist) and procedural fairness (how accusations will be handled, what evidence standards apply, how students can contest charges).

\myparagraph{What should schools NOT do?} We argue that using AI detector outputs as primary or sole evidence for academic integrity cases is (at least partially) responsible for the emotional harm documented here.
Combined with peer-reviewed evidence of detector unreliability~\cite{liang2023,weber2024} and vendor guidance against such use~\cite{turnitin2024}, this likely makes detection-based enforcement both unfair and counterproductive.

\myparagraph{What can schools do?} We call for widespread adoption of process-based assessment approaches, such as staged drafting with revision history, oral defense of written work, in-class writing components, and explicit AI-use policies with disclosure expectations.
These mechanisms can verify authorship without relying on probabilistic tools that produce asymmetric harm~\cite{sullivan2023,redesign2023,cornell2024}.

\subsection{Limitations}\label{sec:limitations}

\myparagraph{Platform Bias.} Naturally, Reddit is not representative of all students or teachers, as users skew younger, more tech-savvy, and more male than the general population~\cite{redditstats}. Results describe discourse patterns within these communities, not the views of all educational stakeholders.

\myparagraph{Group Classification.} While validation suggests 88--94\% accuracy for the subreddit-as-stakeholder proxy, misclassification occurs. This misclassification attenuates observed effects; thus, true stakeholder differences may be larger than reported.

\myparagraph{LLM Annotation.} Labels were generated by an LLM. LLM-human agreement (macro $\kappa = 0.48$--$0.61$) is comparable to human-human agreement ($\kappa = 0.42$), suggesting disagreement reflects task ambiguity rather than LLM-specific error. Whereas, key findings rely on high-agreement labels (AI\_Detector\_Related $\kappa = 0.74$--$0.81$).

\myparagraph{Independence and sample size.} Some users post multiple times, violating independence assumptions for chi-square tests. However, sensitivity analysis with author deduplication shows primary findings are robust.

Also, the student subsample ($n = 204$) is substantially smaller than teachers ($n = 2,769$), reflecting genuine differences in subreddit activity rather than sampling choices. Student-specific estimates have wider confidence intervals; we emphasize the direction of effects and consistency across groups rather than precise student prevalence estimates. The detector$\rightarrow$emotion association is significant within the student subsample ($\chi^2 = 7.19$, $p < 0.01$), but magnitude estimates should be treated as suggestive.

\myparagraph{Longitudinal Studies.} Our dataset covers January 2023 to October 2024. As AI tools and detection systems continue to evolve, findings may not generalize to future periods. %

\subsection{Ethics Considerations}\label{sec:ethics} 
We follow best practices according to the ACM-community code of ethics and professional conduct~\cite{gotterbarn2018acm}.
All data collected and analyzed as part of this study is publicly available and voluntarily posted by the users.
Furthermore, our analysis primarily focuses on aggregate statistics; in a few cases, we report quotes from users to illustrate case-study behaviors, but without revealing their username or including any personally identifiable information.
To ensure reproducibility and comply with the FAIR data principles~\cite{fair}, we will share the source code for our experiments as well as LLM prompts upon request.

\section{Conclusion}
This paper presented a multi-faceted analysis of how different Reddit education-related communities discuss generative AI in high school education.
We focused on learning outcomes, academic integrity and enforcement, AI detection tools, and emotional framing.

Our analysis of 3,789 Reddit posts in five subreddits provides the first large-scale quantitative evidence that AI detector deployment creates measurable emotional harm distributed asymmetrically across educational stakeholders.

Overall, we find that detector-related discussions are associated with more negative emotion across all stakeholder groups, highlighting an emotional cost of AI enforcement beyond questions of detector accuracy.
Teachers tend to frame AI as both potentially helpful and harmful for learning, while students more often discuss it in practical terms centered on enforcement and consequences rather than pedagogy.
The negative emotional impact of detector-related discourse appears stronger for students and mixed communities than for teachers, suggesting the burden falls more heavily on those being monitored than on those enforcing it.

Our analysis complements prior work in this space: while detector unreliability has been documented~\cite{liang2023,weber2024}, our study shows that the \textit{deployment} of these tools---not just their technical limitations---creates emotional harm visible in public discourse. This reframes the detector problem from a measurement issue (``detectors are inaccurate'') to a governance issue (``detector deployment harms students'').

We argue that detection-based enforcement should be abandoned as a primary integrity strategy. The emotional harm, combined with peer-reviewed evidence of detector unreliability~\cite{liang2023,weber2024}, vendor guidance against high-stakes use~\cite{turnitin2024}, and institutional recommendations for alternative approaches~\cite{iowa2024,niu2024}, creates a compelling case for process-based assessment redesign.
Moreover, the broader lesson extends beyond education: weak technical tools deployed under institutional power imbalance create governance failures that harm those with the least power to contest them.

As part of future work we plan to extend this analysis to other online platforms, examine longitudinal changes as AI tools and policies evolve, and evaluate the impact of alternative assessment strategies that reduce reliance on detection-based enforcement.

{\small
\bibliographystyle{apalike}
\bibliography{references}
}

\appendix

\section{LLM Use Statement}

Large language models were used in this research for two purposes: (1) topic relevance filtering of candidate posts, and (2) applying a human-validated annotation scheme to the full dataset. LLMs were \textit{not} used for theory formation, category generation, or interpretation of results. The annotation ontology was developed by human coders; the LLM functioned as a measurement instrument for scaling human-validated coding. All qualitative interpretation and theoretical framing were conducted by human authors. No synthetic text or references were generated. Generative AI tools were used solely for grammar checking and language polishing. All research design, methodology, development, data analysis, and interpretation were performed by the authors.

\section{LLM Annotation Prompts}

\subsection{Annotation Guide Prompt}

The following prompt was used to annotate posts with binary labels. The model was instructed to apply each label independently (TRUE/FALSE), with FALSE as the default when uncertain.

\begin{quote}
You must annotate the post EXACTLY using the following binary labels.
Use TRUE for yes, FALSE for no. If unclear, default to FALSE.
These are NOT mutually exclusive – multiple may apply.

helps\_learning  
Mark this when the post shows AI being used in a way that genuinely supports learning. This includes students using AI to understand material, improve writing, prepare for tests, or get feedback. It also includes teachers using AI to make lesson plans, create classroom materials, customize instruction, or help students learn more effectively. Use this only when the post clearly shows a benefit to learning.

harms\_learning  
Use this when AI is described in a way that makes learning worse. Examples include students relying on AI instead of learning, loss of critical thinking, cheating, plagiarism, or teachers noticing weaker work because of AI. Also use this if the post says AI encourages shortcuts or hurts real skill development.

fear\_anxiety  
Choose this when the post shows worry, nervousness, uncertainty, guilt, or fear about AI. This can be students feeling anxious about their abilities, worrying that AI will replace skills, feeling unsure about rules, or teachers being nervous about AI in the classroom. Focus on emotional tone rather than the topic.

anger\_frustration  
Use this for posts that show annoyance or frustration. This includes people being irritated with students cheating, upset about AI bans, annoyed at changes in learning, or expressing any kind of emotional pushback related to AI. If the tone is sharp or irritated, use this.

excitement\_hope  
Use this when the writer sounds positive or optimistic. Examples include enthusiasm about how AI can help with school work, excitement about new tools, teachers feeling hopeful that AI improves productivity, or students enjoying the benefits. Pick this when the emotional tone leans upbeat or future-oriented.

cheating\_academic\_integrity  
Use this when the post talks about cheating, plagiarism, copying work with AI, getting answers from AI when it is not allowed, or getting in trouble for that. It also includes discussion of rules about what counts as cheating with AI in school.

tutoring\_learning\_support  
Use this when AI is described as a helper for learning, similar to a tutor. This includes explaining concepts, guiding through problems, checking work, giving examples, or helping someone understand a topic better in a legitimate way.

Potential\_to\_help  
Use this when the post talks about how AI could help learning if used correctly, even if it is not helping right now.

AI\_Detector\_Related  
Use this when the post talks about AI detectors or detection systems.

negative\_emotion  
Use this when the post expresses negative emotion of any kind.

positive\_learning\_effect  
Use this when the post clearly indicates that AI improved learning outcomes.
\end{quote}

\subsection{Post Filtering Prompt}

The following prompt was used to determine whether posts were relevant to the research question.

\begin{quote}
You are a binary classifier for Reddit posts.

Research Question: How do Reddit discussions view ChatGPT’s effect on high school (K–12) students’ digital literacy and critical thinking?

Answer ONLY ``YES'' or ``NO''.

Label YES only if ALL are true:
1) Mentions ChatGPT or a named generative AI tool (ChatGPT, GPT-4, Claude, Bard, Copilot, Gemini, etc.).
2) Explicit K–12 context (e.g., high school, middle school, elementary, grade 9–12, K-12 teacher/classroom).
3) The content concerns learning outcomes or academic integrity.

Otherwise label NO.

Examples:
Q: “My high school started banning ChatGPT in exams; students rely on it for essays.” A: YES  
Q: “University professors discuss AI grading tools.” A: NO  
Q: “Why is ChatGPT censoring the name David Mayer?” A: NO  
Q: “Teacher here: students used ChatGPT to write their history essays; I’m seeing less critical analysis.” A: YES
\end{quote}

\end{document}